\documentclass[useAMS,usenatbib]{mn2e}

\usepackage{lmodern}
\usepackage[T1]{fontenc}
\usepackage[pdftex]{graphicx}
\usepackage{aas_macros}
\usepackage{amsmath,amsfonts,amssymb}

\renewcommand{\vec}[1]{\ensuremath{\boldsymbol{#1}}}
\newcommand{\der}[2]{\ensuremath{\frac{\mathrm{d}{#1}}{\mathrm{d}{#2}}}}
\newcommand{\dder}[2]{\ensuremath{\frac{\mathrm{d^2}{#1}}{\mathrm{d}{#2}^2}}}
\newcommand{\pder}[2]{\ensuremath{\frac{\partial{#1}}{\partial{#2}}}}
\newcommand{\rmi}{\ensuremath{\mathrm{i}}}
\newcommand{\rme}{\ensuremath{\mathrm{e}}}
\newcommand{\rmd}{\ensuremath{\mathrm{d}}}
\newcommand{\ph}{\ensuremath{\mathrm{ph}}}
\newcommand{\zaver}[1]{\ensuremath{\left<{#1}\right>_z}}

\title[Excitation of SD waves: WKBJ theory]
{The excitation of spiral density waves through turbulent
fluctuations in accretion discs I: WKBJ theory}
\author[T.~Heinemann and J.~C.~B.~Papaloizou]
{T.~Heinemann$^1$ and J.~C.~B.~Papaloizou$^1$\\
$^1$University of Cambridge, Wilberforce Road, Cambridge CB3 0WA}

\date{\today}

\pagerange{\pageref{firstpage}--\pageref{lastpage}} \pubyear{2008}

\begin{document}

\label{firstpage}

\maketitle

\begin{abstract}
We study and elucidate the mechanism of spiral density wave excitation in a
differentially rotating flow with turbulence which could result from the
magneto-rotational instability. We formulate a set of wave equations with
sources that are only non-zero in the presence of turbulent fluctuations.  We
solve these in a shearing box domain, subject to the boundary conditions of
periodicity in shearing coordinates, using a WKBJ method. It is found that,
for a particular azimuthal wave length, the wave excitation occurs through a
sequence of regularly spaced swings during which the wave changes from leading
to trailing form. This is a generic process that is expected to occur in
shearing discs with turbulence. Trailing waves of equal amplitude propagating
in opposite directions are produced, both of which produce an outward angular
momentum flux that we give expressions for as functions of the disc parameters
and azimuthal wave length.

By solving the wave amplitude equations numerically we justify the WKBJ
approach for a Keplerian rotation law for all parameter regimes of interest.
In order to quantify the wave excitation completely the important wave source
terms need to be specified. Assuming conditions of weak nonlinearity, these
can be identified and are associated with a quantity related to the potential
vorticity, being the only survivors in the linear regime. Under the additional
assumption that the source has a flat power spectrum at long azimuthal wave
lengths, the optimal azimuthal wave length produced is found to be determined
solely by the WKBJ response and is estimated to be $2\pi H$, with $H$ being
the nominal disc scale height. In a following paper by Heinemann \&
Papaloizou, we perform direct three dimensional simulations and compare
results manifesting the wave excitation process and its source with the
assumptions made and the theory developed here in detail, finding excellent
agreement.
\end{abstract}


\section{Introduction}\label{sec:introduction}
Accretion discs are ubiquitous in astrophysics, occurring in close binary
systems, active galactic nuclei and around protostars \citep[see e.g.][for
reviews]{1995ARA&A..33..505P,1996ARA&A..34..703L}. Ever since their importance
was first realized it has been clear that some form of turbulence is necessary
to provide the anomalous angular momentum transport implied by observed
luminosities and inferred accretion rates. This has usually been parametrised
using the \citet{1973A&A....24..337S} $\alpha$-parametrisation.

The most likely source of turbulence is through the
magneto-rotational instability (MRI)
\citep[see][]{1991ApJ...376..214B,1998RvMP...70....1B}. Both local and global
simulations that display sustained MRI turbulence have been performed
with and without net flux. In all cases prolific spiral density
(SD) wave excitation has been noted (e.g.\ \citealt{2005AIPC..784..475G} in the
local case and \citealt{1998ApJ...501L.189A} in the global case). These
waves may be crucial for explaining various phenomena in accretion disk
systems. In the context of protoplanetary disks for instance, it has recently
been suggested that stochastic gravitational forces derived from density
variations due to SD waves may play an important role in driving the migration
of low mass protoplanets \citep{2004MNRAS.350..849N,2005A&A...443.1067N}. This
possibly remains a viable mechanism even in so-called dead zones where the
ionization fraction is too low for MHD turbulence to occur, see the recent
simulations by \citealt*{2007ApJ...670..805O}. In general, SD waves may lead to
significantly enhanced angular momentum transport in magnetically inactive
regions of accretion disks.
It is therefore important to gain an understanding of the processes leading to
the excitation of spiral density waves, how generic the phenomenon is, and how
the wave amplitudes scale with physical parameters.

In order to do this, we assume weakly nonlinear conditions, under which the
important source terms for exciting the SD waves are expected to be
proportional to what we call the pseudo potential vorticity (PPV), which is
equal to the potential vorticity to linear order but differs from it in the
nonlinear regime.
The  potentially important role that PPV plays for the excitation of SD waves
in rotating shear flows has been recognised in earlier work
\citep[see e.g.][]{1997PhRvL..79.3178C}.
These authors found by numerically integrating
the linearised equations of motion of compressible, plane Couette flow that
there exists a linear mode coupling between specified vortical perturbations
and (free) SD waves that leads to efficient excitation of the latter (see also
\citealt{2005A&A...437....9B} for a similar study in the context of accretion
disks). However, the possibility of the generation of vortical perturbations
through the action of turbulence has not yet been fully assessed.

It is the purpose of this paper to develop a mathematically rigorous theory of
the excitation of SD waves within the WKBJ framework. Our approach is inspired
by the work of \citet{2004JAtS...61..211V} who developed an analogous theory
for small amplitude inertia-gravity waves in a local, quasi-geostrophic model
of the Earth's atmosphere. Apart from illuminating the excitation process of
SD waves in rotating shear flows by putting it on a firm mathematical basis,
the theory also enables us to calculate the amplitude of the excited waves and
the associated angular momentum transport explicitly.

In a following paper \cite[][paper II]{2008arXiv0812.2471H} we perform
numerical simulations which to study SD wave excitation in MRI driven
turbulence \citep[see also][]{2005AIPC..784..475G,2006ApJ...653..513S} and
make detailed comparisons to the WKBJ theory presented here. In this scenario,
magnetically dominated turbulent stresses cause vortical perturbations which
then in turn lead to the excitation of the observed SD waves. The role that
turbulent stresses play is thus indirect. In this aspect the excitation
process differs from that considered in Lighthill's theory of aerodynamic
noise generation \citep{1952RSPSA.211..564L}.

The plan of this paper is as follows: In section \ref{sec:shearing-box} we
describe the shearing box model, giving the basic equations and defining the
background shear flow in which the hydromagnetic turbulence responsible for
the SD wave excitation, is generated. In section \ref{sec:wave-equations} we
derive equations describing the excitation of SD waves. These take the form of
linear wave equations with both linear and nonlinear source terms that are
determined by the turbulence.
We focus on waves that are nearly independent of the vertical coordinate which
have been found to dominate in simulations carried out in paper II and which
can be dealt with using a vertical averaging procedure.
We formulate the law of conservation of angular momentum for linear,
non-dissipative SD waves and derive a convenient expression for the average
radial flux which we later use to estimate the angular momentum flux arising
due to the excited waves.

In section \ref{sec:wkbj-theory} we go on to develop the WKBJ theory of wave
excitation. A Fourier analysis is carried out enabling each azimuthal wave
number $k_y$ to be considered separately. The WKBJ theory applies to a
shearing box for which the boundary conditions are the imposition of
periodicity in shearing coordinates and involves a sequence of excitations
uniformly spaced and localized in time, during which the wave swings from
leading to trailing. The wave amplitude and wave action produced in a swing
are calculated from a WKBJ formalism involving the evaluation of integrals
along anti-Stokes lines.

We compare the results derived from asymptotic theory to results obtained by
numerically integrating the ordinary differential equations describing the
evolution of the appropriate Fourier amplitude. We find excellent agreement
between these approaches. This agreement persists under all conditions of
interest. This is in spite of the fact that asymptotic theory formally
requires a parameter depending on the azimuthal wave number to be small.
Finally we discuss our results in Section~\ref{sec:discussion}.

\section{The Shearing Box model}\label{sec:shearing-box}

\subsection{Basic set up and equations}
We consider a conducting gas in the shearing box approximation
\citep{1965MNRAS.130..125G}. A Cartesian coordinate system
$(x,y,z)$ with origin at the centre of the box is
adopted. The system rotates with angular velocity
\mbox{$\vec{\Omega}=\Omega\vec{e}_z$}, with $\vec{e}_z$ being the unit vector
in the $z$-direction. This coincides with the angular velocity of the centre
of the box, taken to be in a circular orbit. In the Keplerian case this is
about a central point mass. The lengths of the sides of
the box in the three coordinate directions are $(L_x,L_y,L_z)$ and vertical
stratification is neglected.

The basic equations are those of MHD for an isothermal gas, i.e.\ the
continuity equation
\begin{displaymath}
\pder{\rho}{t} + \nabla\cdot(\rho\vec{v}) = 0,
\end{displaymath}
the momentum equation
\begin{displaymath}
\pder{(\rho\vec{v})}{t} = -c^2\nabla\rho
- 2\vec{\Omega}\times\rho\vec{v} - \rho\nabla\Phi
+ \nabla\cdot\vec{T}' + \nabla\cdot(2\rho\nu\vec{S})
\end{displaymath}
and the induction equation
\begin{displaymath}
\pder{\vec{B}}{t} =
\nabla\times(\vec{v}\times\vec{B} - \eta\nabla\times\vec{B})
\end{displaymath}
where $\rho$ is the density, the velocity is $\vec{v}=(v_x,v_y,v_z)$, the
isothermal sound speed is $c$, the magnetic field is $\vec{B}=(B_x,B_y,B_z)$,
the nonlinear stress tensor $\vec{T}'$ has components
\begin{equation*}
T'_{ij} = B_i B_j - \delta_{ij}\vec{B}^2/2 - \rho v_i v_j,
\end{equation*}
and $\vec{S}$ is the traceless rate-of-strain tensor whose components are
given by
\begin{equation*}
S_{ij} = \frac{1}{2}\left(\pder{v_i}{x_j} + \pder{v_j}{x_i}
- \frac{2}{3}\delta_{ij}\nabla\cdot\vec{v}\right).
\end{equation*}
The kinematic viscosity is $\nu$, the resistivity is $\eta$,
and the combined gravitational and centrifugal potential is given by
\begin{equation*}
\Phi = -q\Omega^2x^2,
\end{equation*}
where for a Keplerian flow the constant $q=3/2$.

The isothermal MHD equations admit the definition of a characteristic length
scale $H=c/\Omega$, which we will refer to as the nominal disc scale height
even though we have neglected vertical stratification.

\subsection{Equations for deviations from the steady state
and boundary conditions}
The background state is taken to have uniform density $\rho_0$, zero magnetic
field and a linear shear corresponding to the velocity
\begin{displaymath}
\vec{v}_0 = -q\Omega x\vec{e}_y.
\end{displaymath}

As we are interested in wave propagation we work in terms of velocity
deviations from the background shear, \mbox{$\vec{u}=\vec{v}-\vec{v}_0$},
which we use to define the linear momentum density per unit volume
\mbox{$\vec{p}=\rho\vec{u}$}. In terms of $\rho$ and $\vec{p}$ the governing
equations now read
\begin{equation}\label{eq:dev-continuity}
\mathcal{D}\rho + \nabla\cdot\vec{p} = 0,
\end{equation}
\begin{equation}\label{eq:dev-motion}
\mathcal{D}\vec{p} = - c^2\nabla\rho - 2\vec{\Omega}\times\vec{p}
+ q\Omega p_x\vec{e}_y + \nabla\cdot\vec{T}
+ \nabla\cdot(2\rho\nu\vec{S}),
\end{equation}
\begin{equation}\label{eq:dev-induction}
\mathcal{D}\vec{B} =
\nabla\times(\vec{u}\times\vec{B} - \eta\nabla\times\vec{B})
- q\Omega B_x\vec{e}_y
\end{equation}
where the differential operator
\begin{equation}\label{eq:Dshear}
\mathcal{D} = \pder{}{t} - q\Omega x\pder{}{y}
\end{equation}
accounts for advection by the linear shear.
Note that here, the nonlinear stress tensor
\begin{equation}\label{eq:non-linear-stress-tensor}
T_{ij} = B_i B_j - \delta_{ij}\vec{B}^2/2 - \rho u_i u_j
\end{equation}
only contains products of deviations from the background state.

We consider equations (\ref{eq:dev-continuity}) to (\ref{eq:dev-induction})
to be subject to periodic boundary conditions in `Lagrangian' (or `shearing')
coordinates given by
\begin{equation}\label{eq:lagrangian-coordinates}
x' = x,\ y' = y + q\Omega t x,\ z' = z,\ \textrm{and}\ t' = t,
\end{equation}
transformation to which removes the explicit $x$-dependence contained in
(\ref{eq:Dshear}) -- albeit at the expense of an explicit time
dependence\footnote{Without loss of generality we have assumed that the two
coordinate systems coincide at $t=0$.}. Re-expressed in `Eulerian' coordinates
$(x,y,z,t)$ the radial boundary condition for any fluid variable $f$ then
reads
\begin{displaymath}
f(x + L_x,y - q\Omega t L_x,z,t) = f(x,y,z,t)
\end{displaymath}
while the azimuthal and vertical boundary conditions are simply
\begin{displaymath}
f(x,y + L_y,z,t) = f(x,y,z,t)
\end{displaymath}
and
\begin{displaymath}
f(x,y,z + L_z,t) = f(x,y,z,t),
\end{displaymath}
respectively.

\section{Wave equations with sources}\label{sec:wave-equations}

In order to proceed we develop equations for the deviation of the state
variables from their background state in order to obtain wave equations with
sources. These can then be used to study the excitation of SD waves
explicitly.
It is known that SD waves can propagate in a strictly isothermal box with no
dependence on the vertical coordinate \citep[see][]{2007A&A...468....1F} and
we have found that such waves are the ones predominantly excited in our
simulations presented in paper II. We
therefore vertically average equations (\ref{eq:dev-continuity}) and
(\ref{eq:dev-motion}), which then become equations for the vertical averages
of the state variables such as $\vec{p}$ and $\rho,$ in order to describe such
waves. When periodic boundary conditions in $z$ are adopted this can be done
without approximation and it is equivalent to adopting $k_z=0$ when Fourier
transforms are considered.  Proceeding in this way, we denote the vertical
average of a quantity $f$ by use of angle brackets as $\zaver{f}$.

To further simplify the analysis we consider the inviscid limit of the
shearing box equations, which then become
\begin{subequations}\label{eq:first-order-equations}
\begin{gather}
\label{eq:drho}
\mathcal{D}\zaver{\rho} + \partial_x \zaver{p_x}
+ \partial_y \zaver{p_y} = 0 \\
\label{eq:dpx}
\mathcal{D}\zaver{p_x} + c^2\partial_x\zaver{\rho}
- 2\Omega\zaver{p_y} = \zaver{\mathfrak{N}_x} \\
\label{eq:dpy}
\mathcal{D}\zaver{p_y} + c^2\partial_y\zaver{\rho}
+ (2-q)\Omega \zaver{p_x} = \zaver{\mathfrak{N}_y}
\end{gather}
\end{subequations}
where we introduced the short-hand $\vec{\mathfrak{N}}=\nabla\cdot\vec{T}$.
At this point we note that in the zero net flux case considered here, the
magnetic field enters the momentum equation only through the nonlinear stress
tensor (\ref{eq:non-linear-stress-tensor}) so that it will not affect the
description of linear SD waves.

Acting on (\ref{eq:dpy}) with $\mathcal{D}$ and rearranging terms yields
\begin{multline}
\label{eq:wavepy}
(\mathcal{D}^2 - c^2\nabla^2 + \kappa^2)\zaver{p_y} = \\
- c^2\partial_x\zaver{\zeta} + (q-2)\Omega\zaver{\mathfrak{N}_x} +
 \mathcal{D}\zaver{\mathfrak{N}_y}
\end{multline}
where $\kappa^2=2(2-q)\Omega^2$ is the square of the epicyclic frequency and
\begin{displaymath}
\zeta = \partial_x p_y - \partial_y p_x + (q-2)\Omega\rho,
\end{displaymath}
which we shall call the pseudo potential vorticity (PPV). To linear order, the
variation of PPV is equal to the variation of potential vorticity (PV),
\begin{displaymath}
Q = \frac{\partial_x u_y - \partial_y u_x + (2-q)\Omega}{\rho},
\end{displaymath}
\citep[see][]{2005ApJ...626..978J}. Written out explicitly, we have
\begin{displaymath}
\zeta - \zeta_0 = \rho_0^2(Q - Q_0) + \textrm{nonlinear\ terms},
\end{displaymath}
where
\begin{displaymath}
\zeta_0 = (q-2)\Omega\rho_0
\quad\mathrm{and}\quad
Q_0 = \frac{(2-q)\Omega}{\rho_0}
\end{displaymath}
are the steady state background values of PPV and PV, respectively.
For disturbances with \mbox{$k_z=0$} and given a barotropic equation of state,
PV is an exactly conserved quantity whereas PPV varies due to nonlinear
stresses,
\begin{align}
\label{eq:evo-vorticity}
\mathcal{D}\zaver{\zeta} &= \partial_x\zaver{\mathfrak{N}_y} - 
\partial_y\zaver{\mathfrak{N}_x}.
\end{align}

We can form wave equations similar to (\ref{eq:wavepy}) for $\rho$ and $p_x$.
Letting $\mathcal{D}$ act on (\ref{eq:drho}) and (\ref{eq:dpx}) yields
\begin{subequations}
\label{eq:waverhopx}
\begin{multline}
\label{eq:waverho}
(\mathcal{D}^2 - c^2\nabla^2 + \kappa^2)\zaver{\rho}
+ 2q\Omega\partial_y \zaver{p_x} = \\
- 2\Omega\zaver{\zeta} - \partial_x\zaver{\mathfrak{N}_x}
- \partial_y\zaver{\mathfrak{N}_y}
\end{multline}
and
\begin{multline}
\label{eq:wavepx}
(\mathcal{D}^2 - c^2\nabla^2 + \kappa^2)\zaver{p_x}
+ 2q\Omega c^2\partial_y \zaver{\rho} = \\
c^2\partial_y\zaver{\zeta} + 2\Omega\zaver{\mathfrak{N}_y}
+ \mathcal{D}\zaver{\mathfrak{N}_x}.
\end{multline}
\end{subequations}

\subsection{Decomposition into shearing waves}
\label{sec:fourier-decomposition}
In the periodic shearing sheet we may expand all fluid variables in a series
of plane wave solutions
\begin{equation*}
\exp\Big(\rmi k_x' x' + \rmi k_y' y'\Big)
\end{equation*}
in the Lagrangian coordinate frame (\ref{eq:lagrangian-coordinates}). Here,
the radial and azimuthal wave numbers
\begin{equation*}
k_x' = \frac{2\pi n_x}{L_x}\ \textrm{and}\ k_y' = \frac{2\pi n_y}{L_y}
\ \mathrm{with}\ n_x,n_y\in\mathbb{Z}.
\end{equation*}
In terms of Eulerian coordinates, the plane wave expansion for any (vertically
averaged) fluid variable $\zaver{f}$ reads
\begin{displaymath}
\zaver{f}(x,y,t) = \sum_{n_x,n_y}\!
\hat{f}(t)\,\exp\Big[\rmi k_x(t)x + \rmi k_y y\Big]
\end{displaymath}
with a time dependent radial wave number
\begin{equation}\label{eq:k_x(t)}
k_x(t) = k_x' + q\Omega t k_y'.
\end{equation}
and constant azimuthal wave number $k_y = k_y'$.

When viewed from the Eulerian coordinate frame, the radial wave number of
non-axisymmetric disturbances (for which $k_y\ne{}0$) changes linearly in time
due to advection by the linear shear, leading to the notion of sheared
disturbances as originally discussed by Kelvin \citep{thomson_stability_1887}.
In the shearing sheet, non-axisymmetric plane waves are therefore often
referred to as shearing waves.
In an astrophysical context, the usefulness of the concept of sheared
disturbances for understanding accretion disk phenomena was first realised by
\citet{1965MNRAS.130..125G}.

It is customary to classify shearing waves according to whether they are
leading, i.e.\ $k_x(t)/k_y<0$, or trailing, i.e.\ $k_x(t)/k_y>0$. Because the
time dependence of the radial wave number (\ref{eq:k_x(t)}) is such that
$k_x(t)/k_y$ always increases monotonically provided that $q\Omega>0$, every
leading wave will eventually become trailing as time progresses. The change
from leading to trailing is referred to as `swing' and occurs when $k_x(t)=0$.
Different shearing waves swing from leading to trailing at different times,
successive swings being separated, for a given $k_y$, by a fixed
time interval \begin{equation}\label{eq:dtswing}
\delta t_\mathrm{s} = \frac{T_\mathrm{orb}}{q k_y L_x},
\end{equation}
where $T_\mathrm{orb}=2\pi/\Omega$ is the orbital period.

We note that because we are dealing with the Fourier transforms of real
quantities, each Fourier coefficient becomes equal to its complex conjugate
under reflection of the wave number such that
\mbox{$\vec{k}\rightarrow{}-\vec{k}$}. This means that we may without loss of
generality consider only \mbox{$k_y>0$}, and then multiply by a factor of two
after taking the real part of the transforms to obtain physical quantities,
which then accounts for \mbox{$k_y<0$}. Thus from now on we consider only
\mbox{$k_y>0$}.

It is now straightforward to write down the SD wave equations (\ref{eq:wavepy})
and (\ref{eq:waverhopx}) in Fourier space. Substituting
\begin{equation*}
\mathcal{D}\to\rmd/\rmd t,\ \partial_x\to\rmi k_x(t),
\ \textrm{and}\ \partial_y\to\rmi k_y
\end{equation*}
we readily obtain
\begin{subequations}\label{eq:fourier-wave-equations}
\begin{multline}
\dder{\hat{\rho}}{t} + \left[\vec{k}^2(t)c^2 + \kappa^2\right]\hat{\rho}
+ 2q\Omega\rmi k_y\hat{p}_x = \\
- 2\Omega\hat{\zeta} - \rmi k_x(t)\hat{\mathfrak{N}}_x
- \rmi k_y\hat{\mathfrak{N}}_y,
\end{multline}
\begin{multline}
\dder{\hat{p}_x}{t} + \left[\vec{k}^2(t)c^2 + \kappa^2\right]\hat{p}_x
+ 2q\Omega c^2\rmi k_y\hat{\rho} = \\
c^2\rmi k_y\hat{\zeta} + 2\Omega\hat{\mathfrak{N}}_y
+ \der{\hat{\mathfrak{N}}_x}{t},
\end{multline}
and
\begin{equation}
\dder{\hat{p}_y}{t} + \left[\vec{k}^2(t)c^2 + \kappa^2\right]\hat{p}_y =
- c^2\rmi k_x\hat{\zeta} + (q-2)\Omega\hat{\mathfrak{N}}_x
+ \der{\hat{\mathfrak{N}}_y}{t}.
\end{equation}
\end{subequations}

\subsection{Angular momentum flux}\label{sec:wave-action}
Conservation of angular momentum for linear waves in the shearing box follows
from invariance of the system under translations along the azimuthal or
$y$-direction.
Here we note that this actually yields a momentum flux that can be converted
into an angular momentum flux by multiplying by the radius of the centre of
the box.
As the latter quantity does not play any role in the box dynamics we can
conveniently set it to be unity making the momentum and angular momentum
fluxes equivalent.
We introduce the Lagrangian displacement which we define through
\begin{equation}\label{eq:displ}
\mathcal{D}\vec{\xi} = \vec{p}/\rho_0 - q\Omega\xi_x\vec{e}_y
\end{equation}
In terms of $\vec{\xi}$, the density deviations from the background state
are given by
\begin{displaymath}
\delta\rho/\rho_0 + \nabla\cdot\vec{\xi} = 0
\end{displaymath}
and the linearised equations of motion become
\begin{displaymath}
\mathcal{D}^2\vec{\xi} = c^2\nabla\nabla\cdot\vec{\xi}
- 2\vec{\Omega}\times\mathcal{D}\vec{\xi} + 2q\Omega^2\xi_x\vec{e}_x.
\end{displaymath}
These also follow from the requirement that the action
\begin{equation*}
S = \int\!\mathcal{L}(\vec{\xi},\mathcal{D}\vec{\xi},\nabla\cdot\vec{\xi})
\,\rmd^3\vec{x}\,\rmd t,
\end{equation*}
\begin{equation}\label{eq:action}
S = \int\!\mathcal{L}\,\rmd^3\vec{x}\,\rmd t,
\end{equation}
with the Lagrangian density given by
\begin{displaymath}
\mathcal{L} =
\frac{\rho_0}{2}\Bigl[\mathcal{D}\vec{\xi}\cdot\mathcal{D}\vec{\xi}
- c^2(\nabla\cdot\vec{\xi})^2
+ 2(\vec{\Omega}\times\vec{\xi})\cdot\mathcal{D}\vec{\xi}
+ 2q\Omega^2\xi_x^2\Bigr],
\end{displaymath}
and the integral being taken over the box and between two arbitrary points in
time, be stationary with respect to arbitrary variations of the Lagrangian
displacement. The angular momentum conservation law follows from the
invariance of the action (\ref{eq:action}) under infinitesimal translations in
the $y$-direction. The resulting form of Noether's theorem yields
\begin{equation*}
\mathcal{D}\left[\frac{\partial\mathcal{L}}{\partial(\mathcal{D}\vec{\xi})}
\cdot\partial_y\vec{\xi}\right] +
\nabla\cdot\left[\frac{\partial\mathcal{L}}{\partial(\nabla\cdot\vec{\xi})}
\partial_y\vec{\xi} - \mathcal{L}\,\vec{e}_y\right] = 0.
\end{equation*}
We thus define the angular momentum density
\begin{equation}\label{eq:angmom-density}
A = -\frac{\partial\mathcal{L}}{\partial(\mathcal{D}\vec{\xi})}
\cdot\partial_y\vec{\xi} =
-\rho_0(\mathcal{D}\vec{\xi} + \vec{\Omega}\times\vec{\xi})
\cdot\partial_y\vec{\xi}
\end{equation}
and the angular momentum flux
\begin{equation}\label{eq:angmom-flux}
\vec{F} = -\frac{\partial\mathcal{L}}{\partial(\nabla\cdot\vec{\xi})}
\partial_y\vec{\xi} + \mathcal{L}\,\vec{e}_y  =
-c^2\delta\rho\,\partial_y\vec{\xi} + \mathcal{L}\,\vec{e}_y,
\end{equation}
which enables us to write
\begin{equation}\label{eq:angmom-conservation}
\mathcal{D}A + \nabla\cdot\vec{F} = 0.
\end{equation}
Note the minus sign in equations (\ref{eq:angmom-density}) and
(\ref{eq:angmom-flux})  which can be determined from considering the action of
an external force \citep[see also][]{1992ApJ...388..438R}. The angular momentum
conservation law (\ref{eq:angmom-conservation}) may be averaged over $y$ and
$z$ to yield
\begin{displaymath}
\partial_t\langle A\rangle_{yz} + \partial_x\langle F_x\rangle_{yz} = 0.
\end{displaymath}

The angular momentum flux as defined in (\ref{eq:angmom-conservation}) is, at
times, inconvenient to work with because it involves the Lagrangian
displacement $\vec{\xi}$. We can derive a related wave action where the
radial flux only depends on mass density and momentum density as follows. From
(\ref{eq:dpy}) and (\ref{eq:displ}) we have to linear order
\begin{equation}
\label{eq:xi_x-py}
\xi_x = \frac{p_y + c^2\partial_y W}{(q-2)\Omega\rho_0},
\textrm{\ where\ }\mathcal{D}W = \delta\rho.
\end{equation}
Inserting (\ref{eq:xi_x-py}) into (\ref{eq:angmom-flux}) yields after some
straightforward algebra
\begin{multline*}
F_x = \frac{c^2}{(2-q)\Omega\rho_0} \\
\biggl\{\delta\rho\,\partial_y p_y
+ \partial_y\Bigl(c^2\delta\rho\,\partial_y W\Bigr)
- \mathcal{D}\biggl[\frac{c^2}{2}\Bigl(\partial_y W\Bigr)^2\biggr]
\biggr\}.
\end{multline*}
The second and the third term in curly brackets can be absorbed in the
$y$-component of the angular momentum flux and in the angular momentum
density, respectively, giving rise to a new wave action conservation law,
\begin{displaymath}
\mathcal{D}A' + \nabla\cdot\vec{F}' = 0,
\end{displaymath}
where the modified radial angular momentum flux
\begin{displaymath}
F'_x = \frac{c^2\delta\rho\,\partial_y p_y}{(2-q)\Omega\rho_0}.
\end{displaymath}
has the desired property that it does not involve the Lagrangian displacement.
Furthermore, this new flux is equal to the radial component of the angular
momentum flux (\ref{eq:angmom-flux}) after averaging over $y$, $z$, and $t$.
But it should be noted that it in order to establish this equality it has been
assumed that no external forces act in the domain.

We will define a suitable temporal averaging procedure further below in
Section~\ref{sec:angular-momentum-flux}. At this point we note that when
evaluated for a single pair of (complex conjugate) shearing waves, and
averaged over both $y$ and $z$, the two equivalent expressions for the radial
angular momentum flux become
\begin{equation}\label{eq:flux-xi}
\langle F_x\rangle_{yz} = -2 k_y c^2
\mathrm{Im}\Bigl(\hat{\xi}_x^\ast\hat{\rho}\Bigr)
\end{equation}
and
\begin{equation}\label{eq:flux-py}
\langle F'_x\rangle_{yz} = \frac{2 k_y c^2}{(2-q)\Omega\rho_0}
\mathrm{Im}\Bigl(\hat{p}_y^\ast\hat{\rho}\Bigr).
\end{equation}
Here, without loss of generality, we have adopted $k_y>0$ as described in
Section~\ref{sec:fourier-decomposition}.

\section{WKBJ theory of wave excitation}\label{sec:wkbj-theory}
In this section we derive a WKBJ theory of the wave excitation that occurs
during a swing cycle and derive an expression for the wave action produced. We
go on to compare this theory in detail with the results of numerical
integrations of the ordinary differential equations governing the time
dependent evolution of the Fourier transforms of the wave amplitudes.
Excellent agreement is
obtained. In paper II we compare results obtained from the WKBJ theory with
those obtained from MRI simulations.

\subsection{The nature of the source terms}\label{sourcenat}
We first need to establish which of the source terms on the right hand sides
of (\ref{eq:fourier-wave-equations}) are primarily responsible for wave
excitation. Inspection of these equations shows that the source terms are of
two kinds. The first kind is proportional to the transform of PPV and the
second kind is proportional to the nonlinear stress tensor. Only terms of the
first kind remain in the linear regime. Thus under conditions of weak
nonlinearity, we would expect them to dominate. An analysis of the relative
contributions found in direct simulations given in paper II shows that the
contribution of the pseudo potential vorticity related terms is the more
important by an order of magnitude confirming the above idea. It is also shown
that the strength of the wave excitation phenomenon for a swinging wave is
directly correlated with the amplitude of the pseudo potential vorticity
transform at the time of the transition from leading to trailing. Here we
reiterate that although the generation of PPV itself is driven by the
nonlinear stresses, see (\ref{eq:evo-vorticity}), the linear source terms
involving PPV survive at linear order if there is a build up over time under
conditions of weak nonlinearity.

Following on from the above discussion, from now on we retain only source
terms that depend on the pseudo potential vorticity. These are proportional
to $\hat{\zeta}$ which we recall is a Fourier amplitude which is given by
\begin{displaymath}
\hat{\zeta}(t) = \frac{1}{L_x L_y}
\int_0^{L_x}\!\!\!\!\int_0^{L_y}\!
\zaver{\zeta}(x,y,t)\,
\exp\Bigl[\rmi k_x(t) x + \rmi k_y y\Bigr]
\,\rmd x\,\rmd y
\end{displaymath}
From the above expression we note that if $\zaver{\zeta}(x,y,t)$ is an
ultimately smooth function, the Riemann Lebesgue lemma allows us to infer that
the source is negligible as $t\rightarrow\pm\infty$ and so is expected to peak
when the wave swings from leading to trailing. Thus the
wave amplitude excitation process should also be localized around this time.

At this point we recall that in a shearing box the wave excitation process for
a fixed $k_y$ appears as a succession of swings, which as seen from
(\ref{eq:k_x(t)}) are separated by the time interval (\ref{eq:dtswing}). For
the longest possible wave length in the $y$ direction given by the box size
this is $L_y/(2\pi q L_x)$ orbital periods. This is independent of the box
size as long as the aspect ratio is fixed. Although this time interval
formally decreases as $L_x$ increases, because of the periodic symmetry
associated with the shearing box, we expect results to be independent of $L_x$
once this is larger than the radial correlation length associated with the
turbulence, expected to be $\sim H$. Accordingly we might expect the
phenomenon to take a similar form in global simulations
\citep[e.g.][]{2005A&A...443.1067N}.

\subsection {Reduction to three uncoupled second order oscillator equations}
As motivated above we will now neglect any nonlinearities in the problem.
After dropping the nonlinear source terms appearing in the SD wave equations
(\ref{eq:wavepy}) and (\ref{eq:waverhopx}) the evolution of $\rho$, $p_x$, and
$p_y$ in Fourier space is governed by
\begin{subequations}
\label{eq:fourier-waverhopx}
\begin{equation}
\dder{\hat\rho}{t} + \Big(\vec{k}^2 c^2 + \kappa^2\Big)\hat\rho
+ 2q\Omega\rmi k_y\hat p_x = -2\Omega\hat\zeta,
\end{equation}
\begin{equation}
\dder{\hat p_x}{t} + \Big(\vec{k}^2 c^2 + \kappa^2\Big)\hat p_x
+ 2q\Omega c^2\rmi k_y\hat\rho = c^2\rmi k_y\hat\zeta,
\end{equation}
\end{subequations}
and
\begin{equation}
\label{eq:fourier-wavepy}
\dder{\hat p_y}{t} + \Big(\vec{k}^2 c^2 + \kappa^2\Big)\hat p_y
= -c^2\rmi k_x\hat\zeta.
\end{equation}
We note that in the absence of nonlinearities the pseudo potential vorticity,
\begin{equation}\label{eq:PPV-definition}
\hat\zeta = \rmi k_x\hat p_y - \rmi k_y\hat p_x + (q-2)\Omega\hat\rho,
\end{equation}
is conserved exactly, i.e.
\begin{equation}\label{eq:PPV-conservation}
\der{\hat\zeta}{t} = 0.
\end{equation}
The wave equation for $\hat p_y$, given by (\ref{eq:fourier-wavepy}),
therefore decouples from those for $\hat\rho$ and $\hat p_x$, given by
(\ref{eq:fourier-waverhopx}). The latter two equations may be decoupled from
each other as well by introducing the pair
\begin{equation}
\label{eq:ppm-pair}
\hat{p}_\pm =\hat{p}_x\pm\hat{\rho} c,
\end{equation}
for which the linearised wave equations read
\begin{equation}
\label{eq:fourier-waveppm}
\dder{\hat p_\pm}{t}
+ \Big(\vec{k}^2 c^2 + \kappa^2 \pm 2q\Omega\rmi k_y c\Big)\hat p_\pm =
c\Big(\rmi k_y c \mp 2\Omega\Big)\hat\zeta.
\end{equation}
We reiterate that the above equations describe the excitation of density
waves and accordingly it may be confirmed that they are absent in the
incompressible limit for which $c \rightarrow \infty$ while maintaining time
derivatives finite. In this limit, it is readily verified from
(\ref{eq:fourier-waverhopx}) and (\ref{eq:fourier-wavepy}) that the excited
or forced flow considered below simply becomes the incompressible flow
associated with a vorticity distribution that is slowly varying through the
action of nonlinear MHD forces. We note the pseudo potential vorticity
becomes the vorticity in that limit.

We can simplify (\ref{eq:fourier-wavepy}) and (\ref{eq:fourier-waveppm})
further by introducing the dimensionless time variable
\begin{equation}\label{eq:tau}
\tau = \frac{k_x c}{\sqrt{k_y^2 c^2 + \kappa^2}}
\end{equation}
and the dimensionless parameter
\begin{equation}\label{eq:epsilon}
\epsilon = \frac{q\Omega k_y c}{k_y^2 c^2 + \kappa^2}
\end{equation}
in terms of which we have
\begin{subequations}
\label{eq:simplified-wkbj}
\begin{equation}
\label{eq:wkbj-py}
\epsilon^2\dder{\hat{p}_y}{\tau} + (\tau^2 + 1) \hat{p}_y
= -c\left(\frac{\rmi k_x c}{k_y^2 c^2 + \kappa^2}\right)\hat{\zeta}
\end{equation}
and
\begin{equation}
\label{eq:wkbj-ppm}
\epsilon^2\dder{\hat{p}_\pm}{\tau} + (\tau^2 + 1 \pm 2\rmi\epsilon)\hat{p}_\pm
= c\left(\frac{\rmi k_y c \mp 2\Omega}{k_y^2 c^2 + \kappa^2}\right)\hat{\zeta}.
\end{equation}
\end{subequations}

We remark that the homogeneous form of these equations with
${\hat{\zeta}}=0$ can be solved in terms of Parabolic Cylinder functions
\citep[e.g.][]{1987MNRAS.228....1N}. However, we did not find this feature
to be useful in the context of this paper. Rather we use the fact that the
form of the inhomogeneous equations suggests an asymptotic expansion in
$\epsilon$. Formally, such an expansion is valid if $\epsilon\ll{}1$. From
(\ref{eq:epsilon}) we can see that this will be the case both in the high and
the low azimuthal wave number limit given by $k_yc\gg\kappa$ and
$k_yc\ll\kappa$, respectively.

In the following, we will derive asymptotic solutions to
(\ref{eq:simplified-wkbj}) based on the smallness of $\epsilon$. However, we
will demonstrate that these approximative solutions show excellent agreement
with the exact solution obtained from direct numerical integration even in the
worst possible case of $k_y{}c=\kappa$ for which $\epsilon$ attains its
maximum value, \mbox{$\epsilon^\mathrm{max}=q\Omega/2\kappa$}.

\subsection{Slowly varying solutions}\label{sec:slowly-varying}
We consider equations (\ref{eq:simplified-wkbj}) for large $|\tau|$, i.e.\ in
the high radial wave number limit. In this limit the second time derivatives
are significant only for solutions that vary rapidly in $\tau$. Such
solutions are expected when sources are absent and correspond to high
frequency oscillations. This suggests that they can be dropped for solutions
that vary slowly with $\tau$. As a first approximation, we drop the double
time derivative in (\ref{eq:simplified-wkbj}) to obtain
\begin{subequations}
\label{eq:balanced-solutions}
\begin{equation}
\bar{p}_y = -c\left(\frac{\rmi k_x c}
{\vec{k}^2 c^2 + \kappa^2}\right)\hat{\zeta}
\end{equation}
and
\begin{equation}
\bar{p}_\pm = c\left(\frac{\rmi k_y c\mp 2\Omega}
{\vec{k}^2 c^2 + \kappa^2 \pm 2q\Omega\rmi k_y c}\right)
\hat{\zeta},
\end{equation}
\end{subequations}
where we have used (\ref{eq:tau}) and (\ref{eq:epsilon}). These approximative
solutions to inhomogeneous problem are the leading order terms of an
asymptotic series expansion in ascending powers of $\epsilon$. The series
diverges near the time of the swing from leading to trailing, i.e.\ near
$\tau=0$, when the double time derivative in (\ref{eq:simplified-wkbj}) becomes
significant and oscillatory solutions to the homogeneous equation must be
taken into account.

The occurrence of such oscillatory solutions may easily be demonstrated by
direct numerical integration of (\ref{eq:simplified-wkbj}), which is just a set
of linear ordinary differential equations. We start the integration in the far
leading phase, i.e.\ at $\tau=\tau_0$ with $\tau_0$ large and negative. In
this limit the balanced solutions (\ref{eq:balanced-solutions}) hold and may
be used as initial conditions. We note that in doing so we have to be careful
not to violate PPV conservation which we know to be exact in linear theory,
see (\ref{eq:PPV-conservation}). This problem arises on account of the
additional time derivative taken to obtain (\ref{eq:wavepy}) and
(\ref{eq:waverhopx}) from (\ref{eq:first-order-equations}).

Formally, the balanced solutions (\ref{eq:balanced-solutions}) are
reconcilable with PPV conservation only in the limit $\tau\to\pm\infty$. We
thus introduce an error if we use these solutions as initial conditions at
some finite $\tau_0$ and we have to make sure that $\tau_0$ is sufficiently
large so that this error is small. In order to be able to quantify this error
during the course of the integration, we express the PPV $\hat{\zeta}$ in
terms of $\hat{\rho}$ and $\hat{p}_x$, see (\ref{eq:PPV-definition}) together
with (\ref{eq:ppm-pair}), and solve (\ref{eq:simplified-wkbj}) as a system of
three coupled ordinary differential equations. (Alternatively, equations
(\ref{eq:first-order-equations}) could be solved directly). PPV conservation
is not guaranteed in this case, but we find empirically that the numerical
integration conserves PPV arbitrarily well depending on how large $\tau_0$
(and thus the error introduced by using the balanced solutions as initial
conditions) is.

Bearing these general remarks in mind, we now discuss a specific numerical
solution to (\ref{eq:simplified-wkbj}). For this purpose we consider Keplerian
shear, i.e.\ $q=3/2$. Because we would like to determine empirically how well
asymptotic theory works when we are far away from the asymptotic limit
$\epsilon\ll{}1$ we take the worst possible case, i.e.\ a shearing wave with
$k_y=\kappa/c=1/H$ so that $\epsilon=q\Omega/2\kappa=3/4$. For the
determination of the initial conditions from the balanced solutions we assume,
without loss of generality, that $\hat{\zeta}=\Omega\rho_0$ in
(\ref{eq:balanced-solutions}).

\begin{figure}
\begin{center}
\includegraphics[width=\columnwidth]{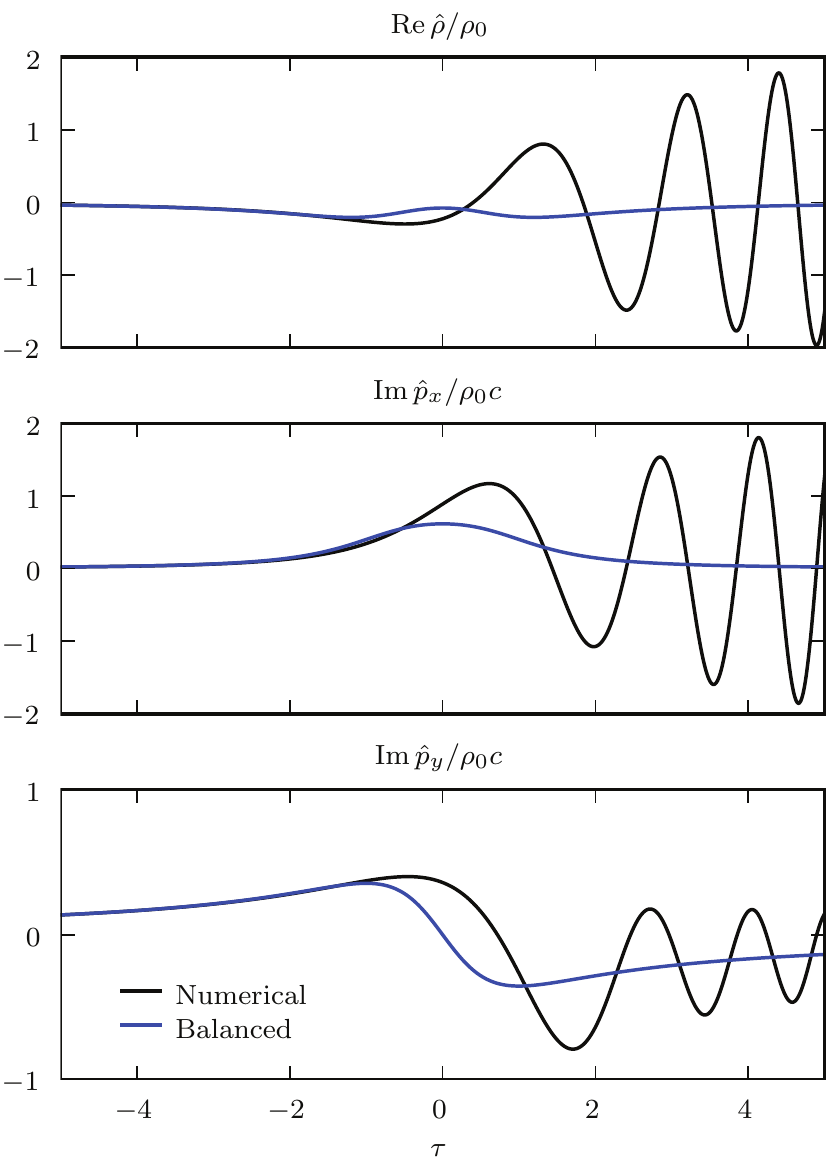}
\end{center}
\caption{Comparison between the numerical solution (black) to the linearised
wave equations (\ref{eq:simplified-wkbj}) with the corresponding balanced
solutions (blue). Here we have used (\ref{eq:ppm-pair}) to compute
$\hat{\rho}$ and $\hat{p}_x$ from $\hat{p}_\pm$. The parameters in this
example are $q=3/2$, $\epsilon=3/4$, and $\hat{\zeta}=\Omega\rho_0$.}
\label{fig:illustrative-balanced}
\end{figure}

We start the integration at $\tau=-100$.\footnote{At this point the relative
error as far as PPV conservation is concerned is $\sim{}10^{-8}$. We find that
the relative error never exceeds $10^{-6}$ during the entire course of the
integration from $\tau=-100$ to $\tau=10$.}
The evolution of the Fourier amplitudes near the swing from leading to
trailing as a function of the radial wave number is shown in
Fig.~\ref{fig:illustrative-balanced}. Before the swing from leading to
trailing the numerical solution closely follows the balanced solution up until
$\tau\approx{}0$ where a sudden transition to oscillatory behaviour occurs. In
the trailing phase, the numerical solution oscillates with a relatively large
but rather slowly evolving amplitude around the balanced solution. Such
oscillatory behaviour cannot be captured by a perturbation series expansion.
In order to describe it we have to resort to singular perturbation theory to
be discussed in the next section.

\subsection{WKBJ solution of the generic oscillator equation}

To study the excitation of a general Fourier mode we seek a solution to the
forced harmonic oscillator equation of the general type
\begin{equation}
\label{eq:forced-oscillator}
\epsilon^2\dder{y(\tau)}{\tau} + (\tau^2 + a^2) y(\tau) = f(\tau)
\end{equation}
with $\epsilon\ll{}1$ and where $a$ is a complex number with
\begin{equation}\label{eq:ph-a}
|a| \sim O(1)
\quad\mathrm{and}\quad
|\ph\,a| < \pi/4,
\end{equation}
the latter being true for $a=\sqrt{1+2\rmi\epsilon}$ with
$0\le\epsilon<\infty$.

Throughout the following analysis make $w = z^{1/n}$ single-valued by taking
the branch cut along the negative real axis and always take the principal
branch, thus
\begin{equation}
\ph\,w = \frac{\ph\,z}{n}.
\label{BRANCH}
\end{equation}

We note that equations (\ref{eq:wkbj-py}) and (\ref{eq:wkbj-ppm}) are special
cases of the above general form. We further also assume that $f(\tau)$ is a
slowly varying function such that $\Delta f\sim O(1)$ for $\Delta\tau \sim
O(1)$ as $\epsilon\to 0$. Physically this means that the pseudo potential
vorticity transform should vary at a slow rate compared to the wave
oscillation frequency at around the time of the swing. This is expected under
conditions of weak nonlinearity as discussed in section \ref{sourcenat} and
simulation results presented in paper II indicate that this is indeed the
case.

\subsection{Outer and balanced solutions}
A solution of equation (\ref{eq:forced-oscillator}) in ascending powers of
$\epsilon^2$ is readily found from regular perturbation theory, and yields to
lowest order
\begin{equation}
y(\tau) = \frac{f(\tau)}{\tau^2 + a^2} + O(\epsilon^2),
\end{equation}
which is, given (\ref{eq:ph-a}), well defined on the entire real axis.
However, this series, giving rise to what is described as the balanced
solution, is only asymptotic. At all
orders in $\epsilon^2$, the series will fail to represent
rapidly changing oscillatory contributions to $y(\tau)$ for which the double
derivative term in (\ref{eq:forced-oscillator}) is significant. These vanish
more rapidly than any power of $\epsilon$ as $\epsilon\rightarrow 0$.

In order to obtain these contributions, that are in fact associated with the
excited SD waves, we have to resort to singular perturbation theory. We thus
seek a solution of the asymptotic form
\begin{align}
y(\tau) &\approx \frac{f(\tau)}{\tau^2 + a^2},
& &\tau \rightarrow -\infty \\
\label{eq:wkbj-ansatz}
y(\tau) &\approx \frac{f(\tau)}{\tau^2 + a^2}
+ \frac{A_{+}\rme^{-\rmi\Phi(\tau)/\epsilon}}{(\tau^2 + a^2)^{1/4}}
+ \frac{A_{-}\rme^{+\rmi\Phi(\tau)/\epsilon}}{(\tau^2 + a^2)^{1/4}},
& &\tau \rightarrow +\infty
\end{align}
which is the sum of the leading order solution to the inhomogeneous problem
obtained from regular perturbation theory, and a standard WKBJ solution to the
homogeneous problem. Here the WKBJ phase is given by
\begin{align}
\Phi(\tau) &= \int_0^\tau\!\!\!\sqrt{\sigma^2+a^2}\,\rmd\sigma \notag\\
&= \frac{1}{2}\left[\tau\sqrt{\tau^2 + a^2}
+ a^2\ln\left(\frac{\tau + \sqrt{\tau^2 + a^2}}{\sqrt{a^2}}\right)\right].
\label{WKBPH}
\end{align}

\subsection{Matching on anti-Stokes lines}

In order to determine the WKBJ (or wave) amplitudes $A_{+}$ and $A_{-}$, we
analytically continue the WKBJ solution (\ref{eq:wkbj-ansatz}) into the
complex $\tau$-plane and match it to inner solutions valid in the immediate
vicinity of the two complex WKBJ turning points at $\tau_\ast = \pm\rmi a$.
Thus in this section the $\pm$ alternative refers to the turning point with
positive and negative imaginary part respectively rather than the
$\hat{p}_{\pm}$ alternative of the previous section.

We find it convenient to match inner and outer solutions along the so-called
anti-Stokes lines defined by
\begin{equation}
\mathrm{Im}\,\Phi(\tau) = \mathrm{Im}\,\Phi(\tau_\ast).
\label{STO}
\end{equation}
Because the imaginary part of the WKBJ phase is constant on these lines, one
of the WKBJ exponentials in (\ref{eq:wkbj-ansatz}) is \mbox{maximally}
sub-dominant to the other one and may thus be ignored.

From the first order Taylor expansion about a turning point we have
\begin{equation}
\label{Taylor}
\tau^2 + a^2 \approx 2\tau_\ast(\tau-\tau_\ast)
\end{equation}
and therefore from equation (\ref{WKBPH}) we obtain
\begin{displaymath}
\Phi(\tau) \approx \Phi(\tau_\ast)
+ \frac{2\sqrt{2}}{3}\tau_\ast^{1/2}(\tau-\tau_\ast)^{3/2}
\end{displaymath}
The condition (\ref{STO}) thus defines three anti-Stokes lines emanating from each
turning point at angles given by
\begin{displaymath}
\ph(\tau-\tau_\ast)
= \left(\mp\frac{5\pi}{6},\,\mp\frac{\pi}{6},\,\pm\frac{\pi}{2}\right)
- \frac{\ph\,a}{3}
\end{displaymath}
For consistency we match on the anti-Stokes lines originating from each turning
point that asymptotically approach the real axis.

\begin{figure}
\begin{center}
\includegraphics[width=\columnwidth]{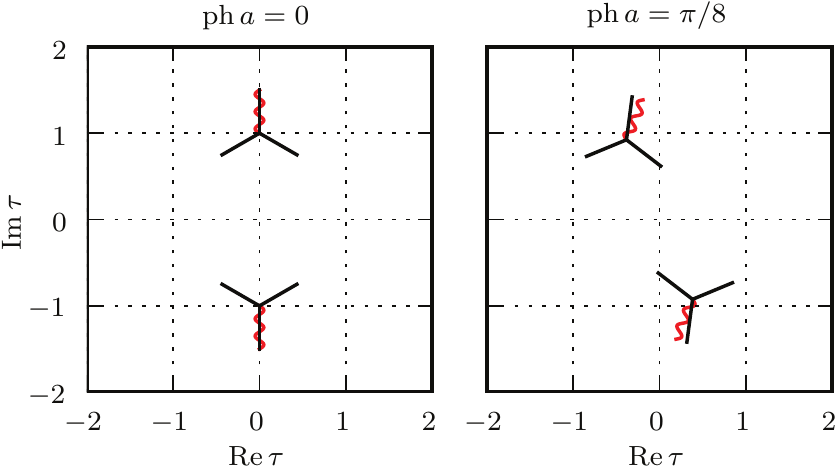}
\end{center}
\caption{Anti-Stokes lines (solid black) emanating from the two turning points
for $\ph\,a=0$ (left panel) and $\ph\,a=\pi/8$ (right panel).
The curly red curves indicate branch cuts.}
\label{fig:wkbj-phase-diagram}
\end{figure}

For $\mathrm{Re}\,\tau < \mathrm{Re}(\pm\rmi a)$ these are
\begin{equation}
\ph(\tau-\tau_\ast) = \mp\frac{5\pi}{6} - \frac{\ph\,a}{3}
\label{STO1}
\end{equation}
and for $\mathrm{Re}\,\tau > \mathrm{Re}(\pm\rmi a)$
\begin{equation}
\ph(\tau-\tau_\ast) = \mp\frac{\pi}{6} - \frac{\ph\,a}{3}
\label{STO2}
\end{equation}
Note that with the definition of the square root obtained from (\ref{BRANCH}),
the WKBJ solution (\ref{eq:wkbj-ansatz}) has branch cuts that leave the
turning points at an angle
\begin{displaymath}
\ph(\tau-\tau_\ast) = \pm\frac{\pi}{2} - \ph\,a
\end{displaymath}
so we can always match along the anti-Stokes lines specified above without
crossing branch cuts. This is illustrated in
Fig.~\ref{fig:wkbj-phase-diagram}.

\begin{figure}
\begin{center}
\includegraphics[width=\columnwidth]{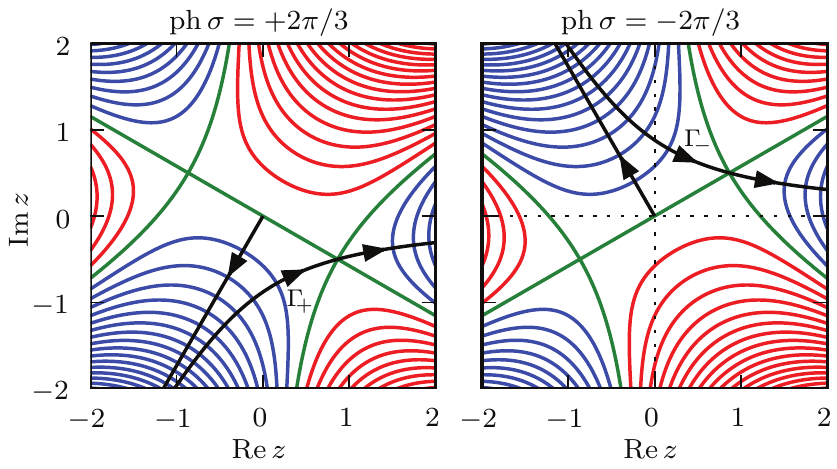}
\end{center}
\caption{Integration contours for the Airy-integral (\ref{eq:airy-int})
in the $z$ plane. In each case the contour is deformed from the positive
real axis to the two paths shown. These contours (black), which extend to
infinity, correspond to the situation where we are on the anti-Stokes lines
that radiate away from the turning points towards the
positive real axis on which $\ph\,\sigma = 2\pi/3$ at the upper turning point (left panel)
and $\ph\,\sigma = -2\pi/3$ at the lower turning point (right panel).}
\label{fig:steepest-descents}
\end{figure}

On an anti-Stokes line one of the two WKBJ exponentials in
$(\ref{eq:wkbj-ansatz})$ will be maximally sub-dominant to the other in the
limit $\epsilon\to{}0$, depending on the sign of the imaginary part of the
WKBJ phase at the turning point from which it emanates. At these we have
\begin{displaymath}
\Phi(\pm\rmi a) = \pm\rmi a^2\pi/4
\end{displaymath}
and thus when the conditions (\ref{eq:ph-a}) are satisfied
\begin{displaymath}
\mathrm{Im}\,\Phi(\pm\rmi a) \gtrless 0.
\end{displaymath}
This means in general that the maximally sub-dominant exponential has an
amplitude smaller by a factor $\propto\exp(-c_0/\epsilon)$, where $c_0$ is a
constant of order unity when compared to the dominant one, which is very small
for small $\epsilon$. Therefore it should be neglected. Dropping the maximally
sub-dominant WKBJ exponential, in the vicinity of the turning points our WKBJ
solution is to leading order in $\tau\mp\rmi a$
\begin{multline*}
y(\tau) \sim \frac{f(\tau_\ast)}{2\tau_\ast(\tau-\tau_\ast)}
+ \frac{A_\pm \rme^{\pi a^2/4\epsilon}}
{[2\tau_\ast(\tau-\tau_\ast)]^{1/4}} \\
\exp\left[\mp\rmi\epsilon^{-1}\frac{2\sqrt{2}}{3}
\tau_\ast^{1/2}(\tau-\tau_\ast)^{3/2}\right]
\end{multline*}
We see that matching on the anti-Stokes line emanating from the upper (lower)
turning point will only enable us to determine the WKBJ amplitude $A_{+}$
($A_{-}$). We will thus need to match on the anti-Stokes lines from both
turning points in order to determine the full WKBJ solution.

\subsection{Inner solution}
To obtain the inner solutions we consider the governing equation
(\ref{eq:forced-oscillator}) in the vicinity of the turning points. Using
the first order Taylor expansion around a turning point (\ref{Taylor}) this
becomes
\begin{equation}\label{eq:forced-oscillator-approx}
\epsilon^2\dder{y(\tau)}{\tau} + 2\tau_\ast(\tau-\tau_\ast)y(\tau)
= f(\tau_\ast).
\end{equation}
We now defined rescaled variables through
\begin{gather}
\label{eq:rescale-sigma}
\sigma = \rme^{\pm 2\pi\rmi/3}(2\tau_\ast)^{1/3}
\epsilon^{-2/3}(\tau-\tau_\ast) \\
\label{eq:rescale-phi}
y(\tau) = \rme^{\pm 2\pi\rmi/3}(2\tau_\ast)^{-2/3}
\epsilon^{-2/3}f(\tau_\ast)\phi(\sigma)
\end{gather}
The first of these indicates that the solutions we seek vary on a scale
$|\tau| \sim \epsilon^{2/3} \ll 1$. This feature enables us to perform an
asymptotic expansion valid for large $|\sigma|$ and still remain in the
vicinity of the turning points. In terms of the rescaled variables, equation
(\ref{eq:forced-oscillator-approx}) yields an inhomogeneous Airy-type equation
\begin{equation}\label{eq:airy-type}
\dder{\phi(\sigma)}{\sigma} + \sigma\phi(\sigma) = 1.
\end{equation}
We remark that from equation (\ref{eq:rescale-sigma}) we deduce that
\begin{displaymath}
\ph\,\sigma = 0
\quad\mathrm{for}\quad
\ph(\tau-\tau_\ast) = \mp\frac{5\pi}{6} - \frac{\ph\,a}{3}
\end{displaymath}
which applies on the anti-Stokes lines given by (\ref{STO1}) and
\begin{displaymath}
\ph\,\sigma = \pm\frac{2\pi}{3}
\quad\mathrm{for}\quad
\ph(\tau-\tau_\ast) = \mp\frac{\pi}{6} - \frac{\ph\,a}{3}
\end{displaymath}
which applies on the anti-Stokes lines given by (\ref{STO2}).

\subsection {Integral representation for the inner solutions}
The solution to (\ref{eq:airy-type}) can be written as an integral
\begin{equation}\label{eq:airy-int}
\phi(\sigma) = \int_0^\infty \rme^{h(\sigma,z)}\rmd z
\quad\mathrm{with}\quad h(\sigma,z) = -\sigma z - z^3/3.
\end{equation}
We are interested in the large-$\sigma$ asymptotic behaviour of this solution,
which is applicable on each of the anti-Stokes lines emanating from both
turning points.

For the anti-Stokes lines corresponding to $\ph\,\sigma=0$ we integrate along
the positive real axis and get an end-point contribution via Watson's lemma
\begin{displaymath}
\phi(\sigma)\sim \frac{1}{\sigma}\quad(\sigma\to\infty,\ \ph\,\sigma = 0)
\end{displaymath}
or with (\ref{eq:rescale-sigma}) and (\ref{eq:rescale-phi})
\begin{displaymath}
y(\tau) \sim \frac{f(\tau_\ast)}{2\tau_\ast(\tau-\tau_\ast)}
\end{displaymath}
which matches our outer solution (\ref{eq:wkbj-ansatz}) to the left of the
turning points if there we set $A_{+} = A_{-} = 0$ consistent with the
causality requirement of the lack of existence of excited waves
for \mbox{$\tau\rightarrow-\infty$}.

For the anti-Stokes lines corresponding to \mbox{$\ph\,\sigma=\pm{}2\pi/3$} we
deform the integration contour from the positive real axis to a contour
consisting of two separate line segments, writing
\begin{equation}\label{eq:airy-int-split}
\int_0^\infty\rme^{h(\sigma,z)}\rmd z
= \int_0^{\infty\,\rme^{\mp 2\rmi\pi/3}}\rme^{h(\sigma,z)}\rmd z
+ \int_{\Gamma_{\!\pm}}\rme^{h(\sigma,z)}\rmd z.
\end{equation}
The first integral on the right hand side is evaluated along a straight line
that goes from the origin to $\infty\,\rme^{\mp 2\pi\rmi/3}$. Along this line,
$h(z)$ is purely real and negative, and we again get an end-point contribution
via Watson's lemma,
\begin{gather*}
\int_0^{\infty\,\rme^{\mp 2\rmi\pi/3}}\rme^{h(z)}\rmd z \sim \frac{1}{\sigma}
\\
\quad(\sigma\to\infty,\ \ph\,\sigma = \pm 2\pi/3).\notag
\end{gather*}
The curve $\Gamma_{\!\pm}$ along which the second integral on the right hand
side of (\ref{eq:airy-int-split}) is taken goes from
$\infty\,\rme^{\mp{}2\pi\rmi/3}$ along a path of steepest descent through the
saddle point $z_\mathrm{s}=\mp\rmi\sigma^{1/2}$ and from there to $\infty$,
see Fig.~\ref{fig:steepest-descents}. In the limit $\sigma\to\infty$, most of
the contribution to this integral will come from near the saddle point
$z_\mathrm{s}$, at which $h(z_\mathrm{s})=\pm(2/3)\rmi\sigma^{3/2}$ and so
\begin{gather*}
\int_{\Gamma_{\!\pm}}\rme^{h(z)}\rmd z
\sim \frac{\pi^{1/2}\rme^{\pm(2/3)\rmi\sigma^{3/2}}}
{\rme^{\mp\pi\rmi/4}\sigma^{1/4}} \\
\quad\left(\sigma\to\infty,\ \ph\,\sigma = \pm 2\pi/3\right).\notag
\end{gather*}
We thus have
\begin{gather*}
\phi(\sigma) \sim \frac{1}{\sigma}
+ \frac{\pi^{1/2}\rme^{\pm(2/3)\rmi\sigma^{3/2}}}
{\rme^{\mp\rmi\pi/4}\sigma^{1/4}} \\
\quad\left(\sigma\to\infty,\ \ph\,\sigma = \pm 2\pi/3\right).\notag
\end{gather*}
Using (\ref{eq:rescale-sigma}) and (\ref{eq:rescale-phi}) we may reexpress the
above solution in terms of $y$ and $\tau -\tau^\ast$ to obtain
\begin{multline*}
y(\tau) \sim \frac{f(\tau_\ast)}{2\tau_\ast(\tau-\tau_\ast)}
+ \frac{\rme^{\pm 3\pi\rmi/4}(\pi/\epsilon)^{1/2}f(\tau_\ast)}
{(2\tau_\ast)^{3/4}(\tau-\tau_\ast)^{1/4}} \\
\exp\left[\mp\rmi\epsilon^{-1}\frac{2\sqrt{2}}{3}
\tau_\ast^{1/2}(\tau-\tau_\ast)^{3/2}\right]
\end{multline*}

We match this solution to our outer solution (\ref{eq:wkbj-ansatz}) in the
neighbourhood of each turning point by equating the factors in front of the
maximally dominant WKBJ exponentials. This then yields the amplitudes
\begin{displaymath}
A_{\pm} = \pm\rmi f(\pm\rmi a)
\left(\frac{\pi}{2a\epsilon}\right)^{1/2}\rme^{-\pi a^2/4\epsilon}.
\end{displaymath}
Using these, the full asymptotic solution takes the form
\begin{multline}\label{eq:full-wkbj}
y(\tau) \approx \frac{f(\tau)}{\tau^2 + a^2}
+ \left(\frac{2\pi}{a\epsilon}\right)^{1/2}
\frac{\rme^{-\pi a^2/4\epsilon}}{(\tau^2 + a^2)^{1/4}} \\
\frac{1}{2\rmi}\left[f(-\rmi a)\,\rme^{+\rmi\Phi(\tau)/\epsilon}
- f(+\rmi a)\,\rme^{-\rmi\Phi(\tau)/\epsilon}\right].
\end{multline}

\subsection{Determination of the wavelike forms of $\hat{p}_y$, $\hat{p}_\pm$,
$\tilde{p}_x$, and $\tilde{\rho}$}

We are now in a position to obtain explicit solutions for $\hat{p}_y$ and
$\hat{p}_\pm$. To bring the wave equation (\ref{eq:wkbj-py}) for $\hat{p}_y$
into the form of the general oscillator equation (\ref{eq:forced-oscillator})
we set
\begin{displaymath}
a = 1
\quad\textrm{and}\quad
f(\tau) = -\frac{\rmi c\tau\hat{\zeta}}{\sqrt{k_y^2 c^2 + \kappa^2}}.
\end{displaymath}
To obtain the wavelike part of the solution, i.e.\ the component proportional
to the WKBJ exponential, which we will denote by a tilde, $f(\tau)$ has to be
evaluated at the complex turning point $\tau_\ast$. Because PPV is conserved
in linear theory this poses no difficulties and we immediately obtain
\begin{equation}\label{eq:pywave-nondim}
\tilde{p}_y = \frac{\rmi\hat{\zeta}_\mathrm{s}c}{\sqrt{k_y^2 c^2 + \kappa^2}}
\sqrt{\frac{2\pi}{\epsilon}}\frac{\rme^{-\pi/4\epsilon}}{(\tau^2 + 1)^{1/4}}
\cos\Bigl[\Phi(\tau)/\epsilon\Bigr],
\end{equation}
where $\hat{\zeta}_\mathrm{s}=\hat{\zeta}(0)$ denotes the PPV amplitude
and the time of the swing and the WKBJ phase
\begin{equation}\label{eq:py-phase}
\Phi(\tau) = \frac{1}{2}\left[\tau\sqrt{\tau^2 + 1}
+ \ln\left(\tau + \sqrt{\tau^2 + 1}\right)\right].
\end{equation}
Using the definitions of $\tau$ and $\epsilon$, respectively given by
(\ref{eq:tau}) and (\ref{eq:epsilon}), we can rewrite the solution for
$\tilde{p}_y$ as
\begin{equation}\label{eq:pywave}
\tilde{p}_y = \rmi H\hat{\zeta}_\mathrm{s}
\mathcal{A}\sqrt{\Omega/\omega}
\cos\left(\int_0^t\!\omega\,\rmd t'\right),
\end{equation}
where the dimensionless amplitude
\begin{equation}\label{eq:wave-amplitude}
\mathcal{A} = \frac{\Omega^{1/2}}{(k_y^2 c^2 + \kappa^2)^{1/4}}
\,\sqrt{\frac{2\pi}{\epsilon}}\,\rme^{-\pi/4\epsilon}
\end{equation}
and we have defined the SD wave frequency
\begin{displaymath}
\omega = \sqrt{\vec{k}^2 c^2 + \kappa^2}.
\end{displaymath}

This procedure is readily repeated to obtain explicit solutions for
$\hat{p}_\pm$. In this case, comparison of the governing equation
(\ref{eq:wkbj-ppm}) with (\ref{eq:forced-oscillator}) implies that we need to
set
\begin{equation*}
a = \sqrt{1\pm 2\rmi\epsilon}
\quad\textrm{and}\quad
f(\tau) =
c\left(\frac{\rmi k_y c\mp 2\Omega}{k_y^2 c^2 + \kappa^2}\right)\hat{\zeta}
\end{equation*}
The solutions for $\tilde{p}_\pm$ determined from (\ref{eq:full-wkbj}) are thus
\begin{multline}\label{eq:ppmwave-nondim}
\tilde{p}_\pm = \rmi\hat{\zeta}_\mathrm{s}c
\left(\frac{2\Omega\mp\rmi k_y c}{k_y^2 c^2 + \kappa^2}\right)
\frac{1}{(1\pm 2\rmi\epsilon)^{1/4}} \\
\sqrt{\frac{2\pi}{\epsilon}}
\frac{\rme^{-\pi/4\epsilon}}{(\tau^2 + 1 \pm 2\rmi\epsilon)^{1/4}}
\sin\Bigl(\Phi_\pm/\epsilon\Bigr),
\end{multline}
where the WKBJ phase
\begin{multline*}
\Phi_\pm = \frac{1}{2}\left[\tau\sqrt{\tau^2 + 1\pm 2\rmi\epsilon}\right. + \\
\left.
(1\pm 2\rmi\epsilon)\ln\left(\frac{\tau + \sqrt{\tau^2 + 1\pm 2\rmi\epsilon}}
{\sqrt{1\pm 2\rmi\epsilon}}\right)\right].
\end{multline*}
It is important to note here that $\Phi_\pm$ are not real with the consequence
that quantities such as $\rme^{\rmi\Phi_{+}/\epsilon}$ have a power law as
well as exponential dependence on $\tau$ for large $\tau$.

Again, we rewrite (\ref{eq:ppmwave-nondim}) in more familiar terms using
(\ref{eq:tau}) and (\ref{eq:epsilon}) to obtain
\begin{equation}\label{eq:ppmwave}
\tilde{p}_\pm = \rmi\hat{\zeta}_\mathrm{s}H
\mathcal{A}_\pm\sqrt{\Omega/\omega_\pm}
\sin\left(\int_0^t\!\omega_\pm\,\rmd t'\right),
\end{equation}
where this time the amplitude is given by
\begin{multline}\label{eq:Aplus}
\mathcal{A}_\pm =
\frac{\Omega^{1/2}}{(k_y^2 c^2 + \kappa^2 \pm 2q\Omega\rmi k_y c)^{1/4}} \\
\left(\frac{2\Omega\mp\rmi k_y c}{\sqrt{k_y^2 c^2 + \kappa^2}}\right)
\sqrt{\frac{2\pi}{\epsilon}}\,\rme^{-\pi/4\epsilon}
\end{multline}
and the wave frequency is
\begin{displaymath}
\omega_\pm = \sqrt{\omega^2 \pm 2q\Omega\rmi k_y c}.
\end{displaymath}
Note that both the amplitude and the wave frequency are complex valued.

Recalling that \mbox{$\hat{p}_\pm=\hat{p}_x\pm\hat{\rho} c$}, it is now a
simple matter to determine the wavelike contributions to $\hat{p}_x$ and
$\hat{\rho}$ in the forms 
\begin{subequations}
\begin{equation}\label{eq:rhowave}
\tilde{\rho} = -\hat{\zeta}_\mathrm{s}\Omega^{-1}
\,\mathrm{Im}\left[\mathcal{A}_{+}
\sqrt{\Omega/\omega_{+}}\sin\left(\int_0^t\!\omega_{+}\,\rmd t'\right)\right]
\end{equation}
and
\begin{equation}\label{eq:pxwave}
\tilde{p}_x = \rmi\hat{\zeta}_\mathrm{s} H
\,\mathrm{Re}\left[\mathcal{A}_{+}
\sqrt{\Omega/\omega_{+}}\sin\left(\int_0^t\!\omega_{+}\,\rmd t'\right)\right]
\end{equation}
\end{subequations}
where we have used the fact that \mbox{$\mathcal{A}_{+}=\mathcal{A}_{-}^\ast$}
and \mbox{$\omega_{+}=\omega_{-}^\ast$}.

\begin{figure}
\begin{center}
\includegraphics[width=\columnwidth]{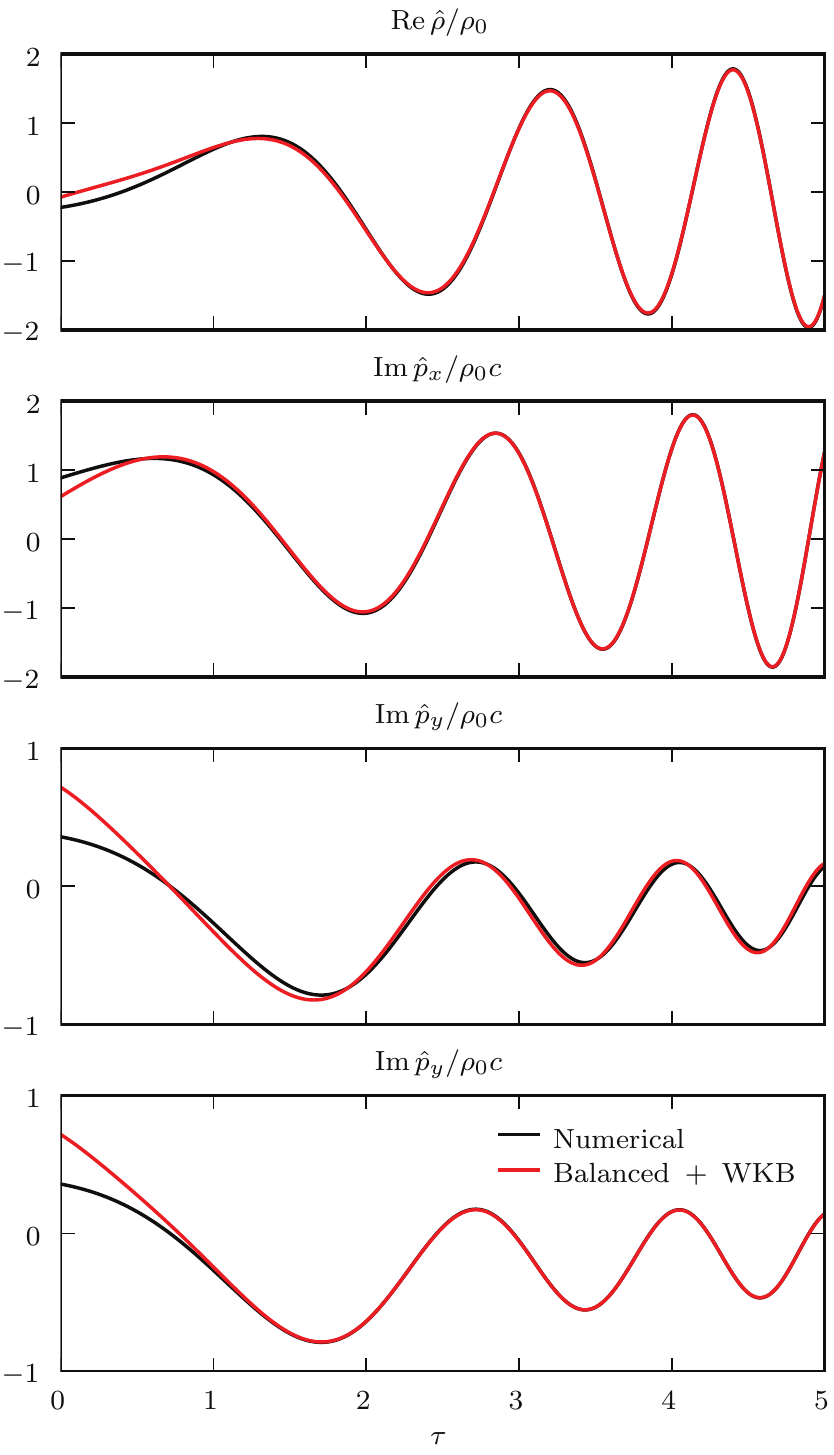}
\end{center}
\caption{Comparison between the numerical solution (black) to the linearised
wave equations (\ref{eq:simplified-wkbj}) with the full asymptotic, i.e.\
balanced plus WKBJ, solution (red). The parameters are the same as in
Fig.~\ref{fig:illustrative-balanced}, i.e.\ $q=3/2$, $\epsilon=3/4$, and
$\hat{\zeta}=\Omega\rho_0$. In the lowermost panel, $\hat{p}_y$ was obtained
from (\ref{eq:pywave-alt}), while in the panel immediately above it was
obtained from (\ref{eq:pywave}).}
\label{fig:illustrative}
\end{figure}

We have seen in Section~\ref{sec:slowly-varying} that the balanced solutions
(\ref{eq:balanced-solutions}) show good agreement with the exact numerical
solution in the leading phase but fail to capture the oscillatory behaviour in
the trailing phase. Having derived the wave like WKBJ solutions of the
homogeneous wave equations, i.e.\ (\ref{eq:pywave}), (\ref{eq:rhowave}) and
(\ref{eq:pxwave}), we are now in a position to determine whether they
correctly describe this oscillatory behaviour.

In Fig.~\ref{fig:illustrative} we compare the exact numerical solution
discussed in
Section~\ref{sec:slowly-varying} with the full asymptotic solution, i.e.\ the
sum of the balanced and the WKBJ solution, in the trailing phase ($\tau>0$).
There is excellent agreement
between the full asymptotic solutions and the
numerical solutions.  Beyond $\tau=2$ the asymptotic
solutions are virtually indistinguishable from the numerical solutions.
Remarkably, this is so even though for this example we have chosen a shearing
wave with an intermediate azimuthal wave number of $k_y{}c=\kappa$ for which
the `small' parameter $\epsilon$ attains its maximum value, $\epsilon=3/4$ in
the case of Keplerian shear considered here, and we are therefore as far away
as possible from the asymptotic limit $\epsilon\ll{}1$.

We note that in the case of $\hat{p}_y$ there is a small but noticable
discrepancy between the numerical solution and the full asymptotic solution
obtained directly from (\ref{eq:pywave}). However, we can derive an
alternative expression for $\hat{p}_y$ from PPV conservation. Because the WKBJ
solutions are solutions to the free wave equations they should carry no PPV,
from which it follows that
\begin{equation}\label{eq:pywave-alt}
\tilde{p}_y = \frac{\rmi k_y\tilde{p}_x + (2-q)\Omega\tilde{\rho}}{\rmi k_x}
\end{equation}
This expression agrees with (\ref{eq:pywave}) in the limit $\epsilon\ll{}1$
and we see from Fig.~\ref{fig:illustrative} that (\ref{eq:pywave-alt}) is in
fact more accurate for $\epsilon\sim{}1$.

We comment that after we have reintroduced the spatial dependence by
multiplying with \mbox{$\exp(\rmi\vec{k}\cdot\vec{x})$} and then taking the
real part, these solutions are found to consist of two waves of equal
amplitude travelling in opposite directions. This is a natural outcome given
the symmetries of the shearing box. However, for the same reason, both waves
transport angular momentum in the same direction, i.e.\ outward if they are
trailing, see Section~\ref{sec:angular-momentum-flux}.

\subsection{Asymptotic behaviour of the WKBJ solutions}
The WKBJ solutions for $\tilde{\rho}$ and $\tilde{p}_x$ given by
(\ref{eq:rhowave}) and (\ref{eq:pxwave}), respectively, involve complex phases
which disguises their large time asymptotic behaviour. To make this more
apparent we note that for large times or equivalently $\tau$ we have
\begin{displaymath}
\Phi_{+} \sim \Phi + \rmi\epsilon\left[\frac{1}{2} + \ln(2\tau)\right]
- \frac{(1+2\rmi\epsilon)\ln(1+2\rmi\epsilon)}{4},
\end{displaymath}
where $\Phi$ is given by equation (\ref{eq:py-phase}) and is purely real. We
see that the imaginary part of $\Phi_{+}$ will result in an extra power of
$\tau$ when taking the exponential. This has the consequence that when the
sine of the WKBJ phase in (\ref{eq:rhowave}) and (\ref{eq:pxwave}) is
re-expressed in terms of exponentials only those with absolute values that
increase with $\tau$ need to be retained. In this case these are
\mbox{$\propto\exp(-\rmi\Phi_{+}/\epsilon)$} and their asymptotic form is
given by
\begin{multline*}
\exp\Bigl(-\rmi\Phi_{+}/\epsilon\Bigr) \sim
2\tau\exp\Bigl(-\rmi\Phi/\epsilon\Bigr) \\
\frac{1}{\sqrt{1 + 2\rmi\epsilon}}
\exp\left[\frac{1}{2}
- \frac{\ln(1+2\rmi\epsilon)}{4\rmi\epsilon}\right],
\end{multline*}
from which it follows that
\begin{displaymath}
\sin\left(\int_0^t\!\omega_{+}\,\rmd t'\right) \sim
\rmi\left(\frac{\omega}{\Omega}\right)
\exp\left(-\rmi\int_0^t\!\omega\,\rmd t'\right)
\frac{\mathcal{B}_{+}}{\mathcal{A}_{+}},
\end{displaymath}
where we have defined
\begin{equation}\label{eq:Bplus}
\mathcal{B}_{+} =
\frac{\Omega\,\mathcal{A}_{+}}{\sqrt{k_y^2 c^2+\kappa^2 + 2q\Omega\rmi k_y c}}
\exp\left[\frac{1}{2} - \frac{\ln(1+2\rmi\epsilon)}{4\rmi\epsilon}\right].
\end{equation}
Using the above relations we can readily find the large $\tau$ asymptotic
forms of $\tilde{\rho}$ and $\tilde{p}_x$ given by
\begin{subequations}
\begin{equation}\label{eq:rho-large-tau}
\tilde{\rho} \sim
-\hat{\zeta}_\mathrm{s}\Omega^{-1}|\mathcal{B}_{+}|\sqrt{\omega/\Omega}\,
\cos\left(\int_0^t\!\omega\,\rmd t' - \mathrm{ph}\,\mathcal{B}_{+}\right)
\end{equation}
and
\begin{equation}\label{eq:px-large-tau}
\tilde{p}_x \sim
\rmi\hat{\zeta}_\mathrm{s}H|\mathcal{B}_{+}|\sqrt{\omega/\Omega}\,
\sin\left(\int_0^t\!\omega\,\rmd t' - \mathrm{ph}\,\mathcal{B}_{+}\right).
\end{equation}
\end{subequations}
We thus see that because the WKBJ amplitudes for $\tilde{\rho}$ and
$\tilde{p}_x$ are complex valued, see (\ref{eq:Aplus}), the envelope of the
oscillation grows in time and there is non-trivial phase shift with respect to
$\tilde{p}_y$.

\subsection{The angular momentum flux}\label{sec:angular-momentum-flux}
In section \ref{sec:wave-action} we obtained two equivalent expressions for
the radial angular momentum flux $F_x$ which uses the radial Lagrangian
displacement and $F'_x$ which uses the $y$-component of the momentum density.
We determined the angular momentum flux for a single pair of (complex
conjugate) shearing waves. Here, we are interested in the angular momentum
flux associated with the excited waves, i.e.\ with the WKBJ solutions. In
(\ref{eq:flux-xi}) and (\ref{eq:flux-py}) we therefore replace $\hat{\rho}$,
$\hat{p}_y$, $\hat{\xi}_x$ by $\tilde{\rho}$, $\tilde{p}_y$, $\tilde{\xi}_x$,
respectively, and obtain
\begin{equation}\label{eq:flux-xi-tilde}
\langle F_x\rangle_{yz} = -2 k_y c^2
\mathrm{Im}\Bigl(\tilde{\xi}_x^\ast\tilde{\rho}\Bigr)
\end{equation}
and
\begin{equation}\label{eq:flux-py-tilde}
\langle F'_x\rangle_{yz} = \frac{2 k_y c^2}{(2-q)\Omega\rho_0}
\mathrm{Im}\Bigl(\tilde{p}_y^\ast\tilde{\rho}\Bigr),
\end{equation}
where $k_y>0$ is understood.

We can use the WKBJ solutions just obtained to determine these angular
momentum fluxes. In order to calculate (\ref{eq:flux-xi-tilde}) we need an
explicit expression for the radial Lagrangian displacement which in Fourier
space is given by
\begin{displaymath}
\hat{\xi}_x = \frac{1}{\rho_0}\int\!\hat{p}_x\,\rmd t.
\end{displaymath}
the leading order WKBJ solution in the limit of large $\tau$ can be calculated
from (\ref{eq:px-large-tau}) directly and is given by
\begin{displaymath}
\tilde{\xi}_x \sim
-\frac{\rmi\hat{\zeta}_\mathrm{s}H}{\Omega\rho_0}\sqrt{\Omega/\omega}
\,|\mathcal{B}_{+}|
\cos\left(\int_0^t\!\omega\,\rmd t' - \mathrm{ph}\,\mathcal{B}_{+}\right)
\end{displaymath}
Using this result together with (\ref{eq:rho-large-tau}) we readily obtain
\begin{displaymath}
\lim_{\tau\to\infty}\overline{\langle F_x\rangle}_{yz} =
\frac{k_y c|\hat{\zeta}_\mathrm{s}|^2 H^2|\mathcal{B}_{+}|^2}{\Omega\rho_0}.
\end{displaymath}
With the help of (\ref{eq:Aplus}) and (\ref{eq:Bplus}) this may be expressed
in the form
\begin{equation}\label{eq:flux-limit}
\lim_{\tau\to\infty}\overline{\langle F_x\rangle}_{yz} =
\frac{2\pi|\hat{\zeta}_\mathrm{s}|^2 H^2}{q\rho_0}
\frac{\Omega(k_y^2 c^2 + 4\Omega^2)\,\rme^{f(\epsilon)}\,\rme^{-\pi/2\epsilon}}
{\left[(k_y^2 c^2 + \kappa^2)^2 + (2q\Omega k_y c)^2\right]^{3/4}},
\end{equation}
where the overline denotes an average over one oscillation period, and we have
defined
\begin{displaymath}\label{eq:f(eps)}
f(\epsilon) = 1 - \frac{\tan^{-1}(2\epsilon)}{2\epsilon}.
\end{displaymath}

Alternatively, we may use equation (\ref{eq:flux-py-tilde}) to determine the
wave action. This involves $\tilde{p}_y$ instead of $\tilde{\xi}_x$ and is
accordingly easier to work with in an Eulerian formulation. The calculation of
(\ref{eq:flux-py-tilde}) is cumbersome if we use the `direct' WKBJ solution
for $\hat{p}_y$, given by (\ref{eq:pywave}), because its non-trivially phase
shifted with respect to the WKBJ solutions for $\hat{\rho}$ and $\hat{p}_x$,
given by (\ref{eq:rhowave}) and (\ref{eq:pxwave}), making the temporal average
over one oscillation period somewhat ill defined. However, if we exploit PPV
conservation and express $\hat{p}_y$ in terms of $\hat{\rho}$ and $\hat{p}_x$,
see (\ref{eq:pywave-alt}), which is also found to be more accurate for
$\epsilon\sim{}1$ (see Fig.~\ref{fig:illustrative}), the calculation is
trivial and we obtain
\begin{equation}
\lim_{\tau\to\infty}\overline{\langle F'_x\rangle}_{yz} =
\lim_{\tau\to\infty}\overline{\langle F_x\rangle}_{yz}.
\end{equation}

\begin{figure}
\begin{center}
\includegraphics[width=\columnwidth]{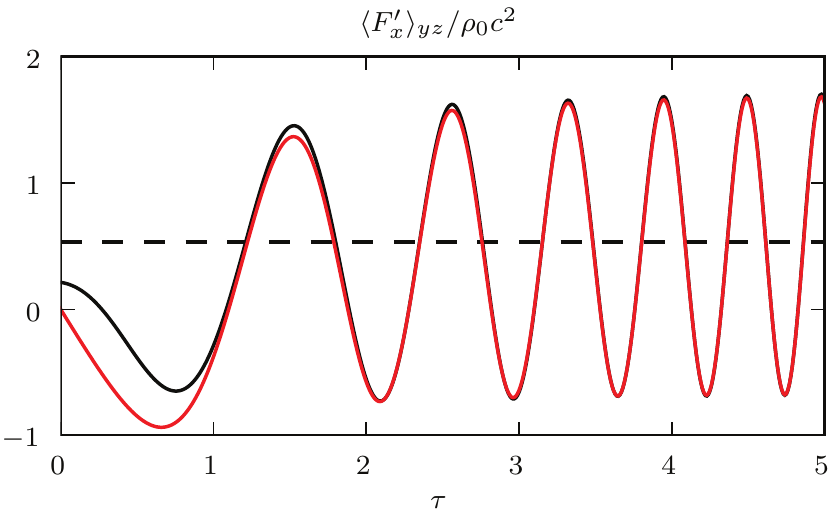}
\end{center}
\caption{Wave action for the wave shown in Fig.~\ref{fig:illustrative}. The
black and red line are the wave action for the wave part of the numerical and
WKBJ solution, respectively. The dashed line indicates the large-$\tau$
average value (\ref{eq:flux-limit}).}
\label{fig:illustrative-wave-action}
\end{figure}

In Fig.~\ref{fig:illustrative-wave-action} we show the angular momentum flux
$\langle F'_x\rangle$ associated with the wave parts of solutions plotted in
Fig.~\ref{fig:illustrative}. In order to obtain the wave part of the numerical
solution we have simply subtracted the balanced solution.

As expected, the WKBJ solutions agrees with the numerical solution remarkably
well. Due to interference between the forward and the backwards travelling
waves actually obtained in our numerical solution, the quantity defined by
(\ref{eq:flux-py-tilde}) oscillates. However, the average over one oscillation
period approaches a constant non-zero value as is expected for a linear wave.

\section{Discussion}\label{sec:discussion}
In this paper we have developed a theory of SD wave excitation in a
rotating shear flow with turbulence which may result
from the magneto-rotational instability but under the assumption that the
magnetic field is too weak to affect the form of the waves significantly. We
considered the commonly adopted shearing box model for which the flow is
subject to the boundary condition of periodicity in shearing coordinates
\citep{1965MNRAS.130..125G}.

The main feature resulting from the shear is that wave excitation occurs
through a sequence of regularly spaced swings as the wave changes from leading
to trailing form. For a fixed azimuthal wave number $k_y$, and a Keplerian
rotation profile, the swings are separated by a time interval
\mbox{$\delta{}t_\mathrm{s}=T_\mathrm{orb}/(qk_yL_x)$}, where
the orbital period $T_\mathrm{orb}=2\pi/\Omega$. For the optimal azimuthal
wave number \mbox{$k_y=\kappa/c$} (see
discussion below) and \mbox{$L_x=H$} as an estimated radial correlation length
of the turbulence, which should also be the minimum box size required to
capture its essential properties, it follows that
\mbox{$\delta{}t_\mathrm{s}\approx{}T_\mathrm{orb}$}.

The wave equations governing the excitation during a particular swing were
found to depend on time alone and under the assumption that the important
source terms causing the wave excitation are associated with the pseudo
potential vorticity, they could be solved to find the asymptotic wave form and
net positive wave action produced. The form of the wave equations necessitated
a WKBJ analysis in the complex
plane.  In this respect the formalism differs from shearing box analyses that
adopt rigid or free boundary conditions, or which assume strictly harmonic
forcing with radial boundaries extended to infinity, rather than periodicity
in shearing coordinates. In the former cases one can separate out a harmonic
time dependence and solve a problem in space for the wave amplitude
\citep[e.g.][]{1987MNRAS.228....1N}.

The analysis of the wave excitation process driven by pseudo potential
vorticity carried out in this paper has similarities to an analysis of
inertia-gravity waves excited in the earth's atmosphere by
\citet{2004JAtS...61..211V} who perform an analogous WKBJ analysis in
the complex plane.

The excitation process produces waves of equal amplitude propagating in
opposite directions. As these waves are both trailing, by symmetry each
produces an equal outward angular momentum flow. Even when the excitation
process is linear, as the waves propagate away, the radial wave length
shortens until shock dissipation
eventually occurs \citep[e.g.][]{2001ApJ...552..793G}. Thus waves are always
likely to be seen to manifest nonlinear effects as the
characteristic radial wave length shortens. When waves behave linearly
during the initial excitation, but subsequently undergo significant but
not complete dissipation between successive swings, the rate of angular
momentum transport can be estimated as the wave action produced in single
swing given by equation (\ref{eq:flux-limit}). This situation is found to
occur in the simulations presented in paper II.

\begin{figure}
\begin{center}
\includegraphics[width=\columnwidth]{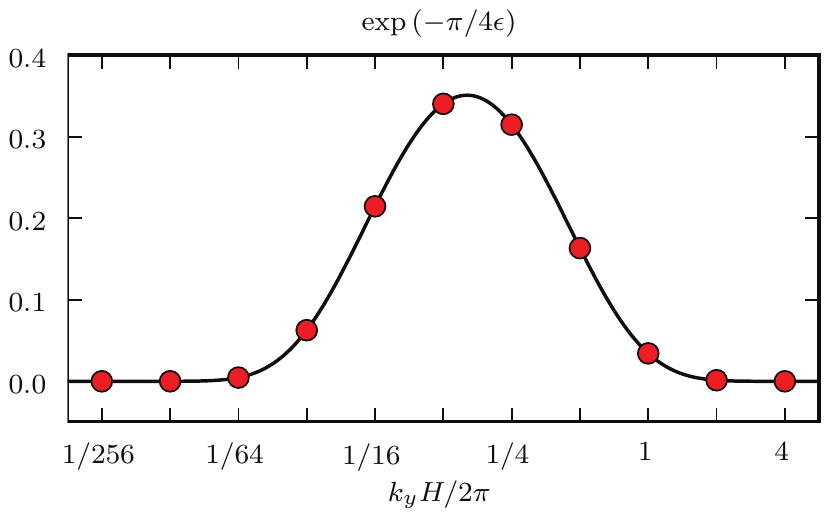}
\end{center}
\caption{Exponential dependence of the WKBJ amplitudes on the azimuthal wave
number $k_y$ for a Keplerian rotation law.}
\label{fig:F-of-eps}
\end{figure}
An important parameter is the value of the
azimuthal/horizontal wave number, $k_y,$ for which the wave excitation is most
favoured, which we define as being the value for which the
wave amplitude produced is maximal.
According to (\ref{eq:pywave}) and (\ref{eq:ppmwave}) the wave amplitude
is a product of a known function of $k_y$ and the square of the
Fourier amplitude of PPV at the time of
swing. The latter quantity, being determined by the nonlinear hydromagnetic
turbulence, cannot be found from the wave excitation calculations
performed here. This aspect is discussed in paper II where relevant numerical
simulations are performed and analysed. Here we shall anticipate results and
assume that the PPV spectrum is relatively flat at small $k_y$ with the
consequence that it does not affect the dependence of the wave amplitude
on $k_y$ significantly.

The wave amplitude depends on the azimuthal wave number $k_y$ through the
parameter
\begin{equation*}
\epsilon = \frac{q\Omega k_y c}{k_y^2 c^2 + \kappa^2}
\end{equation*}
in such a way that it is exponentially small for $\epsilon\ll{}1$, see
(\ref{eq:wave-amplitude}) and (\ref{eq:Aplus}). Given the fact that $\epsilon$
is small both in the small and the long azimuthal wave number limit, we deduce
that wave excitation will be most effective near the optimal wave number
\begin{displaymath}
k^\mathrm{opt}_y = \kappa/c
\end{displaymath}
for which $\epsilon$ takes its maximum value
\begin{displaymath}
\epsilon_\mathrm{max} = q\Omega/(2\kappa).
\end{displaymath}
For a Keplerian disc with $q=3/2$ and $\kappa=1$ we have
\mbox{$k_y^\mathrm{opt}=1/H$} and thus $\epsilon_\mathrm{max}=3/4$. We recall
that WKBJ theory gives very accurate results for values of $\epsilon$ as large
as this. For illustrative purposes we plot the exponential dependence of the
wave amplitudes as a function of azimuthal wave number in
Fig.~\ref{fig:F-of-eps}.  We see that amplitude of the excited wave falls off
rapidly away from the optimal wave number.

The arguments given above suggest that SD wave production will be most
effective for \mbox{$k_y\sim{}k_y^{\rm{opt}}$}. This is the longest possible
azimuthal wave length for a box with \mbox{$L_y=2\pi{}H$} as is commonly
adopted. For boxes of this size and smaller wave production is expected to be
most effective at the longest azimuthal wave length. On the other hand once
$L_y$ exceeds $2\pi{}H$ the longest wave length is expected to no longer be the
most effective. This is fully supported by the simulation results presented
in paper II. In this paper we confirm the main features of the excitation
process described here and verify the dominance of the pseudo potential
vorticity related source terms. Although the waves are observed to become
nonlinear very soon after the initial excitation, the main
features of the analysis presented here are confirmed. This suggests that
useful extensions can be made to the analysis of wave
excitation under more general conditions such as those that incorporate
significant self-gravity. We plan to undertake these in the near future.

\section*{Acknowledgements}
T.~H.\ acknowledges support from the STFC and the Isaac Newton Trust. The
authors wish to thank Stephen J.\ Cowley and James C.\ McWilliams for
rewarding discussions.

\bibliographystyle{mn2e}
\bibliography{theory}

\label{lastpage}

\end{document}